\documentclass[aps, prb, twocolumn, amsmath, amssymb, showpacs,superscriptaddress]{revtex4-1}

\usepackage{graphicx}
\usepackage{times}
\usepackage[pdftex, colorlinks=true, linkcolor=blue, citecolor=blue, urlcolor=blue]{hyperref}
\usepackage{units}

\graphicspath{{./}}
\DeclareGraphicsExtensions{.pdf}

\newcommand{\strasbourg}{Institut de Physique et Chimie des Mat\'{e}riaux 
                         de Strasbourg, 
                         Universit\'{e} de Strasbourg, 
                         CNRS UMR 7504, 23 rue du Loess, BP 43, 
                         F-67034 Strasbourg, France}
\newcommand{\zurich}{Solid State Physics Laboratory, 
                     ETH Z\"{u}rich, CH-8093 Z\"{u}rich, 
                     Switzerland}
\newcommand{\manchester}{School of Physics and Astronomy, 
                         University of Manchester, Oxford Road, 
                         Manchester, M13 9PL, UK}

\begin{document}

\title{Classical origin of conductance oscillations in an integrable cavity} 

\author{Christina P\"{o}ltl}
\affiliation{\strasbourg }

\author{Aleksey Kozikov}
\altaffiliation{Present address: \manchester .}
\affiliation{\zurich }

\author{Klaus Ensslin}
\affiliation{\zurich }

\author{Thomas Ihn}
\affiliation{\zurich }

\author{Rodolfo A. Jalabert}
\affiliation{\strasbourg }

\author{Christian Reichl}
\affiliation{\zurich }

\author{Werner Wegscheider}
\affiliation{\zurich }

\author{Dietmar Weinmann}
\affiliation{\strasbourg }


\begin{abstract}
Scanning gate microscopy measurements in a circular ballistic 
cavity with a tip placed near its center yield a non-monotonic 
dependence of the conductance on the tip voltage. 
Detailed numerical quantum calculations reproduce these conductance 
oscillations, and a classical scheme leads to its physical 
understanding. 
The large-amplitude conductance oscillations are shown to be of 
classical origin, and well described by the effect of a 
particular class of short trajectories. 
\end{abstract}


\maketitle


\section{Introduction}

The Scanning Gate Microscopy (SGM) technique 
\cite{topinka2000, topinka2001, leroy2005, heller2005, jura2009,jura2010,schnez2011a,sellier2011} 
has been developed and applied to study two-dimensional electron 
gases (2DEG) surrounding a Quantum Point Contact (QPC)
and other mesoscopic systems.\cite{schnez2011a,martins2007}
Initially, the goal was to obtain information beyond that provided through  
standard quantum transport experiments by measuring the effect of a 
local potential on the sample conductance. 
However, the interpretation of the SGM measurements remains challenging, 
\cite{jalabert2010, sellier2011, gorini2013, schnez2011a}
in particular because most experiments operate in a regime where 
the potential induced by the SGM tip strongly perturbs the 2DEG.
Performing SGM with tunable geometries indicated that the presence of
confinement considerably affects the interpretation of the data. 
\cite{steinacher16,kozikov16}

A more recent purpose of SGM is the usage of the tip to control and 
modify the potential landscape, and thus the sample geometry,
allowing systematic studies of the effect of sample shape on 
coherent electron transport. 
For the example of a circular cavity connected to leads by QPCs, 
\cite{kozikov2013} the tip has been used to control and study 
magnetoelectric subbands. Another example along these lines is the 
electronic analog of the Braess paradox, \cite{pala2012,sousa2013}
where the tip is used to cut one out of several routes of electron transport
through the sample. 

While the signatures of an underlying classically chaotic electron dynamics 
have been clearly established, \cite{jalabert_scholarpedia}
the situation with integrable geometries is less conclusive due to the lack of 
global stability of the dynamics and the unavoidable effect of smooth disorder 
in the samples.
In particular, the conductance fluctuations and weak localization in circular
cavities have been experimentally studied and compared with other geometries.
\cite{marcus92,berry94,persson95,chang94,lee97}
Intriguingly, the observed behavior did not always correspond to what was 
expected for an integrable geometry.
Controlling the potential landscape of a ballistic cavity with an SGM tip 
allows one to alter the underlying classical dynamics within a given sample, 
and thus to compare different classical dynamics within 
the same sample.

In this paper, we present SGM measurements on a circular cavity exhibiting 
an unexpected non-monotonic dependence of the conductance through the cavity
on the strength of a tip placed over the center.
Detailed numerical modeling of the measured structure reproduced these  
large-amplitude conductance oscillations with tip strength. 
In order to understand the physical mechanism at the origin of the 
conductance oscillations with tip strength, we develop a semiclassical 
approach which demonstrates the key role played by the modification of 
classical trajectories induced by the tip potential. The statistical analysis 
of the ensemble of classical trajectories leads to a detailed understanding 
of the underlying mechanism and the crucial role played by the smoothness of 
the tip potential.

In Sec.\ \ref{sec:experiment} we present the SGM measurements on a 
circular cavity and in particular the non-monotonic dependence of 
the conductance on tip strength when the tip is placed near the center. 
The numerical simulation of the conductance as a function of the 
strength of a tip placed in the cavity center is shown 
in Sec.\ \ref{sec:simulation} for a realistic model that succeeds to 
provide a quantitative description of the measured conductance. 
In Sec.\ \ref{sec:model} we present a simplified model possessing 
nevertheless the essential ingredients to yield the conductance 
oscillations, in qualitative agreement with those of the experiment.
Section \ref{sec:trajectories} discusses the evaluation of the conductance 
based on classical trajectories.
The statistical analysis of the ensemble of trajectories in 
Sec.\ \ref{sec:mechanism} provides a classical understanding of 
the conductance oscillations, and a simplified treatment presented in 
Sec.\ \ref{sec:simpleanalysis} highlights the basic mechanism behind the 
effect. In Sec.\ \ref{sec:off-center} we compare experimental results 
beyond the case of a centered tip with the numerics and a classical 
estimate for the size of the region where the conductance varies 
non-monotonically with tip strength.

\begin{figure*}
\centerline{\includegraphics[width=0.95\textwidth]{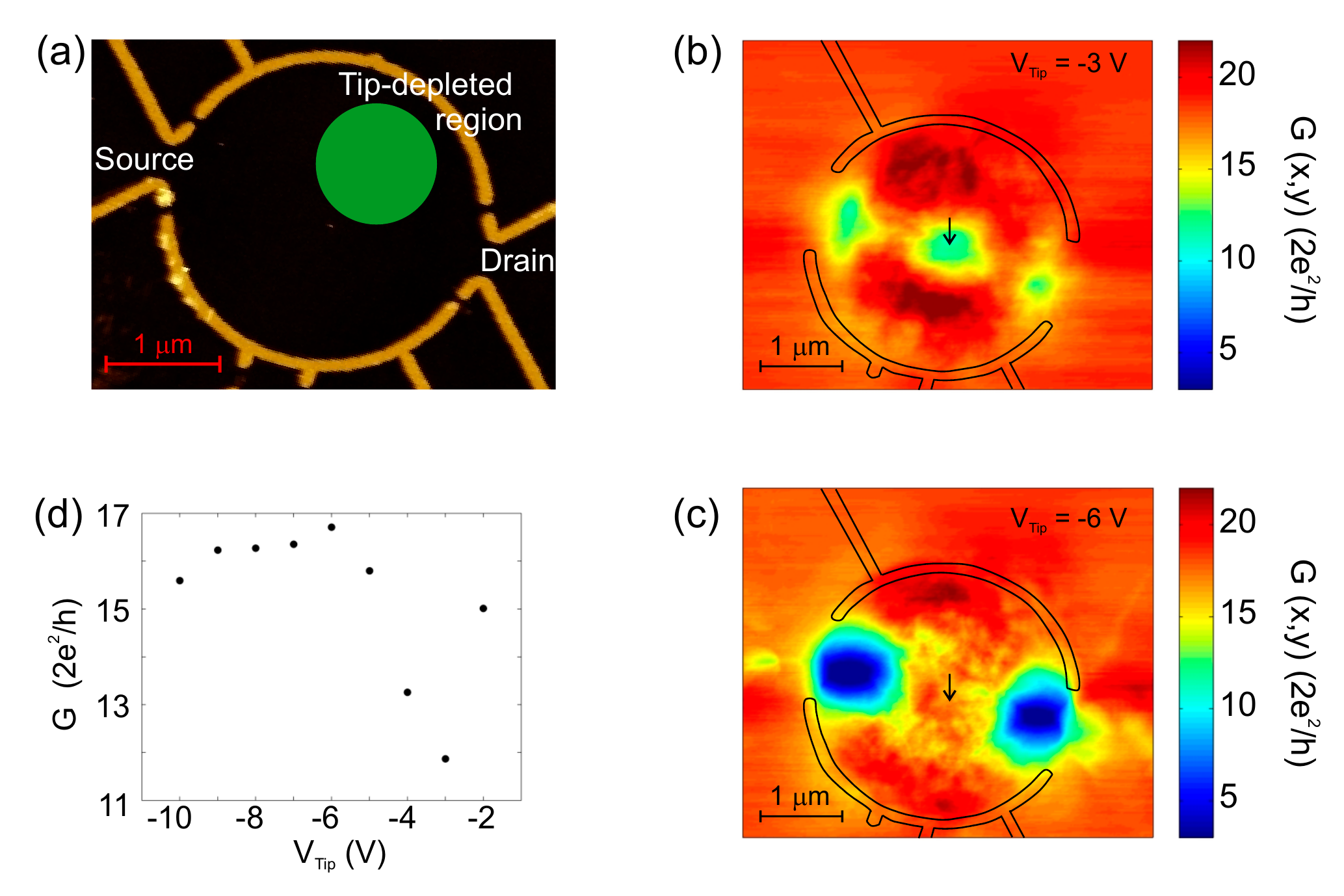}}
\caption{\label{fig:experiment}
(Color online) (a) AFM image of the cavity used in the experiment. 
The black area corresponds to the surface of GaAs. Yellow 
contacts are top gates. 
The tip-induced depleted region is indicated by the green
circle for $V_\mathrm{Tip}\approx \unit[-8]{V}$. 
The current flows between source and drain contacts. (b)-(c) 
Conductance $G$ 
as a function of tip position at a tip voltage of $-3$ and 
\unit[$-6$]{V}, 
respectively. Black lines outline biased top gates (the grounded 
gates are not shown). 
(d) Tip voltage dependence of the conductance when the tip is 
at the center of the cavity [marked by an arrow in (b) and (c)].}
\end{figure*}
\section{Experimental observation}\label{sec:experiment}

The SGM response has been measured in a circular ballistic cavity 
with a 
diameter of about \unit[3]{$\mu$m}, electrostatically defined in a 
GaAs-GaAlAs heterostructure. 
The chosen setup is such that the cavity is connected to source and 
drain by wide openings having a width of about \unit[1]{$\mu$m}. 
The 2DEG is \unit[120]{nm} below the surface, with a 
transport mean free path of \unit[49]{$\mu$m}. 
The Fermi energy is $E_\mathrm{F}=\unit[4.3]{meV}$ and
the Fermi wavelength is about $\lambda_\mathrm{F}=\unit[72]{nm}$.
The size of the structure, being much smaller than the 
elastic mean free path, and the low temperatures used in the experiment, 
set the present study in the coherent ballistic regime.

Figure \ref{fig:experiment} (a) shows an AFM image of the sample where 
top gates (yellow) are placed on the surface of the structure. 
The effect of the SGM tip is controlled by the tip voltage $V_\mathrm{Tip}$.
For $V_\mathrm{Tip} \lesssim \unit[-3.5]{V}$ the tip potential creates a 
depletion disk in the 2DEG whose size increases with increasingly negative 
tip voltages.
The green circle indicates the approximate size of the depletion disk at 
$V_\mathrm{Tip} \approx  \unit[-8]{V}$. 
Figures \ref{fig:experiment} (b) and (c) show the conductance through the 
cavity as a function of tip position for fixed values of $V_\mathrm{Tip}$.
The circular cavity is defined by sufficiently negative voltages applied to 
the top gates indicated by the black lines. The other gates visible in (a) 
are grounded, resulting in a very open cavity.
When the tip is placed close to the entrances of the cavity, there is a 
suppression of the conductance that becomes more pronounced for 
more negative tip voltages.    

A special situation arises when the tip is close to the center. For the less 
negative tip voltage (b), the conductance is suppressed by the effect of the 
tip, while for the more negative $V_\mathrm{Tip}$ (c), we observe an 
enhancement of the conductance. 
Figure \ref{fig:experiment} (d) shows the conductance as a function of 
$V_\mathrm{Tip}$ with the position of the tip fixed at the center 
of the cavity. The non-monotonic dependence of the conductance on the tip 
strength is a surprising observation: \textit{Going for a more invasive configuration, 
with larger depletion disks and further blocking of the area for electron 
transport may result in an enhancement of the conductance!\cite{differentfrombraess}} 

In order to achieve a theoretical understanding of the intriguing behavior 
of the conductance observed in the experiment (Fig.\ \ref{fig:experiment}), 
we perform numerical simulations of models of different complexity and 
develop a semiclassical approach, focusing on the situation where 
the tip is in the center of the cavity.

\begin{figure}
\centerline{\includegraphics[width=\linewidth]{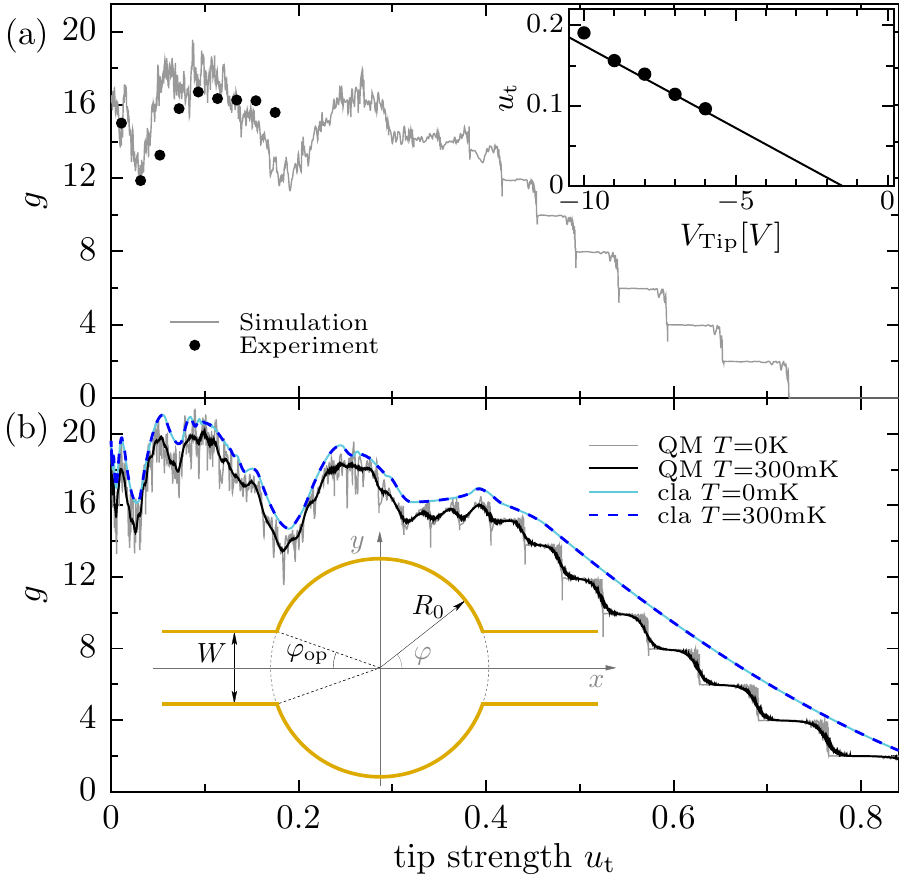}}
\caption{\label{fig:conductance}
(Color online) Dimensionless conductance as a function of the tip strength $u_\mathrm{t}$ 
for (a) the realistic model described in Sec.\ \ref{sec:simulation}
and (b) the simplified model treated in Sec.\ \ref{sec:model}. 
Grey solid lines are for the coherent zero-temperature conductance.
In (a), the dots correspond to the experimental points of 
Fig.\ \ref{fig:experiment} (d) using the linear identification between 
$V_\mathrm{Tip}$ and $u_\mathrm{t}$ shown in the inset of (a) and 
discussed in the text.
In (b), 
the black line represents the quantum conductance at $T=\unit[300]{mK}$. 
The light blue and dark blue (dashed) lines are the classical results 
at $T=\unit[0]{}$ and $T=\unit[300]{mK}$, respectively.
Inset: Sketch of the considered setup. 
A circular cavity with radius $R_0$ is 
connected to quasi-one-dimensional leads of width $W$, $\varphi$ measures the 
angles from the origin and $2\varphi_\mathrm{op}$ is the opening angle of the 
contacts seen from the center of the cavity.}
\end{figure}
\section{Numerical simulation of the experiment}\label{sec:simulation}

Attempting a quantitative description of electronic transport through a 
microstructure is a challenging task, due to the unknown features of the 
self-consistent electronic potential at the basis of one-particle 
modeling.

We start by considering a disorder-free
circular cavity with a tip in its center, using parameters and 
conditions that are as close as possible to the ones of the real sample.
The electrostatic confinement potential of the cavity due 
to the charged top gates is calculated using COMSOL with the exact 
geometry parameters of the sample shown in Fig.\ \ref{fig:experiment}.
We take the standard approach \cite{shawthesis} of modeling 
the SGM tip by a Lorentzian potential 
\begin{equation}\label{eq:lorentzian_potential}
 U_\mathrm{S}(\vec{r}) = 
 \frac{u_\mathrm{t} A}{w_\mathrm{t}^2 + [\vec{r}-\vec{r}_\mathrm{T}]^2}. 
\end{equation}
The constant $A=E_\mathrm{F} R_0^2$ is introduced in order to 
work with a dimensionless tip strength parameter $u_\mathrm{t}$. We choose  
$E_\mathrm{F}$ as the energy-scale and the radius of the cavity 
$R_0=\unit[1500]{nm}$ as the length-scale. \cite{tipnotdependentonR}
The width $w_\mathrm{t}=\unit[200]{nm}$ leads 
to a realistic tip size, of the order of the tip-2DEG distance.
In most of our analysis, the tip position $\vec{r}_\mathrm{T}$ is fixed at 
the center of the cavity.

The coherent zero-temperature conductance was computed with KWANT, 
\cite{kwant-paper} a package that implements a recursive Green 
function algorithm. 
A tight-binding square lattice is used with lattice parameter 
$a=\unit[5]{nm}$, much smaller than $\lambda_\mathrm{F}$. 
The dimensionless conductance $g=G/(2e^2/h)$ is presented and 
discussed henceforth.       
The corresponding numerical results are shown in 
Fig.\ \ref{fig:conductance} (a). 
Superposed to the small-scale ballistic conductance fluctuations,
one observes a large-scale oscillation of the conductance with 
tip strengths for weak and moderately strong tips. 
Such a behavior reproduces the experimentally measured 
non-monotonic SGM response. 

In order to relate the tip strength parameter 
$u_\mathrm{t}$ appearing in the model potential 
\eqref{eq:lorentzian_potential} with the tip voltage $V_\mathrm{Tip}$
used to control the tip strength in the experiment, we follow the procedure 
described in Ref.\  \onlinecite{steinacher15} to estimate 
the size of the depletion disk from the SGM response when the tip is
close to the cavity edges or a QPC.   
Choosing the value of $u_\mathrm{t}$ such that the depletion disk of the 
Lorentzian potential \eqref{eq:lorentzian_potential} corresponds to the  
estimated size, we find the
approximately linear relation between the tip strength
$u_\mathrm{t}$ and the voltage $V_\mathrm{Tip}$ shown in the inset 
of Fig.\ \ref{fig:conductance} (a). We extrapolate the relation 
down to weak tip potentials where no depletion occurs. 
Using such a rough estimate, 
we are able to convert the experimental data of 
Fig.\ \ref{fig:experiment} (d) in order to present them 
in Fig.\ \ref{fig:conductance} (a)
as a function of tip strength $u_\mathrm{t}$, 
together with the numerically obtained conductance.
Besides small deviations in tip strength that may be due to uncertainties 
in the precise shape of the tip potential, the agreement is good, 
confirming the quantitative validity of our model for 
the description of the experimental setup.

It can be noticed that the second conductance maximum, obtained in the 
simulations around $u_\mathrm{t}=0.27$, is beyond the available 
experimental data.
In the regime of very strong tip potentials the conductance exhibits 
plateaus at conductance values that are multiples of $2 \times (2e^2/h)$, 
which decrease with increasing tip strength. 
Such a behavior occurs in the regime where the depletion disc 
generated by the tip becomes so large that the device is reminiscent of 
two parallel quantum wires having the same quantized conductance.

\begin{figure}
\centerline{\includegraphics[width=\linewidth]{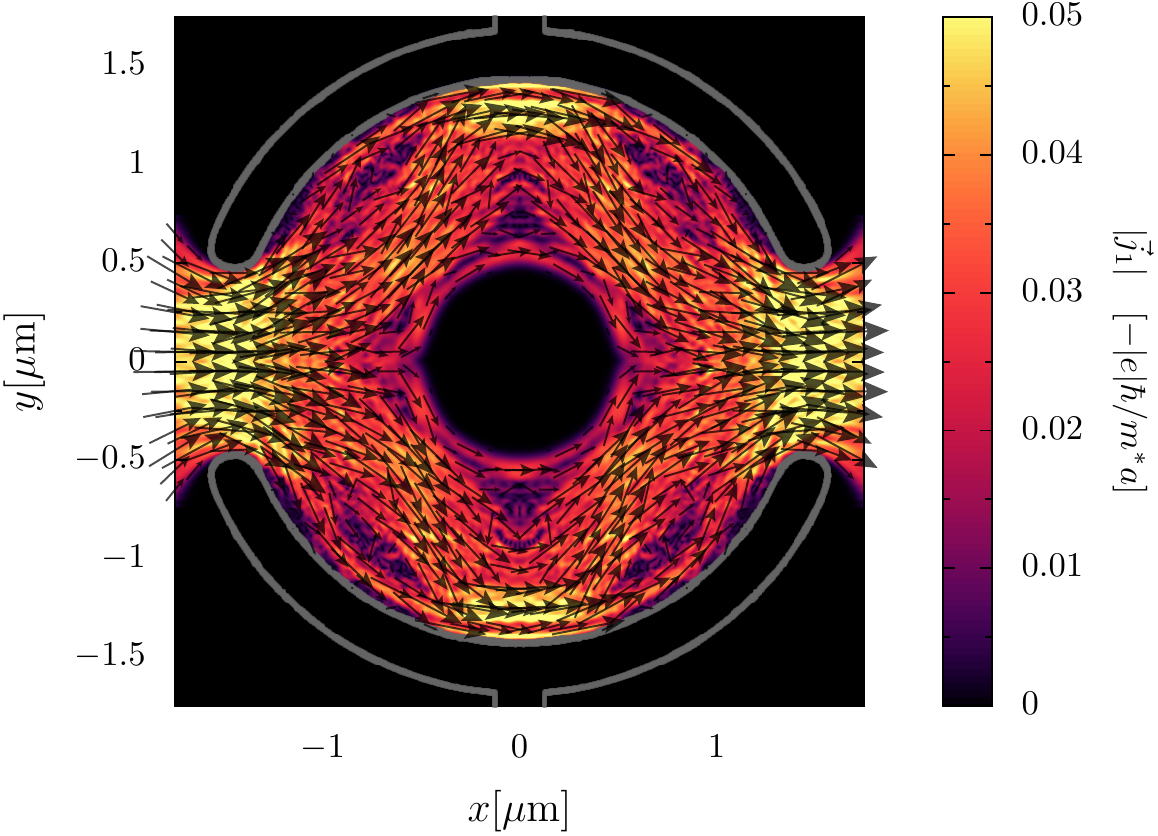}}
\caption{\label{fig:localcurrent}
(Color online)
Current density of scattering states impinging from the left 
through the realistic cavity of 
Fig.\ \ref{fig:conductance} (a), with a tip of strength $u_\mathrm{t}=0.128$
at the center. The color scale gives the absolute value of the 
local current, and thin black arrows reflect the current direction. 
The thick grey lines indicate the region where the confinement 
potential depletes the 2DEG.
}
\end{figure}
The current density is a local quantity of major interest in SGM 
studies. \cite{leroy2005,gorini2013} 
In the invasive regime of the experiments, the current density in the 
cavity strongly depends on the tip strength.
In Fig.\ \ref{fig:localcurrent}, the current density calculated from scattering states at the Fermi energy that are impinging from the left lead 
is shown as a function of the position within the cavity, 
for a tip placed in the center with strength 
$u_\mathrm{t}=0.128$ (close to the first maximum of the 
large-scale oscillations seen in Fig.\ \ref{fig:conductance} (a)). The central 
(black) area of vanishing current density reflects the tip-depleted area.
The diamond-like pattern observed in the current flow around the tip suggests  
the signature of classical electron trajectories following these lines.     

The agreement obtained between the experimental measurements and the 
numerical results in the realistic system (together with the current density 
plots like that of Fig.\ \ref{fig:localcurrent}) constitutes a detailed level of description of quantum transport, but it doesn't allow by itself to 
reach an understanding of the physical mechanism behind the observed 
conductance oscillations. The next step towards such a goal,
made in the following section, consists in 
elaborating a simplified model that exhibits conductance oscillations 
and, at the same time, allows for the identification and analysis of the 
classical trajectories. 

\section{Conductance oscillation in a simplified model system}
\label{sec:model}

In order to understand the mechanisms underlying the non-monotonic 
tip-dependence of the conductance and to find the key ingredients 
for the occurrence of the phenomenon, we should attempt to reduce the 
complexity of the model. 
Though it is tempting to assume hard wall boundaries for the cavity 
and for the tip potential, no significant large-scale conductance 
oscillations occur within such a model. Our analysis in 
Sec.\ \ref{sec:simpleanalysis} provides an explanation for the failure of a
fully hard-wall model to yield conductance oscillations. 

We found that a hard wall cavity attached to 
semi-infinite leads as sketched in the inset of 
Fig.\ \ref{fig:conductance} (b), perturbed by a tip-potential of the form 
\begin{equation}\label{pot_minimal}
U_\mathrm{M}(\vec{r})= \frac{u_\mathrm{t} A}{(\vec{r}-\vec{r}_\mathrm{T})^2}
\end{equation}
is a simplified, minimal model that reproduces all the features observed in 
the realistic simulation. \cite{failureofothermodels}
An important advantage of this model is that the classical trajectories can 
be found analytically. The parameter $A$ is defined after 
Eq.\ \eqref{eq:lorentzian_potential}. In this model we took
$R_0=\unit[1500]{nm}$ as the radius of the cavity and $W=\unit[1000]{nm}$ 
for the width of the leads. 

Figure \ref{fig:conductance} (b) shows the conductance as a function of tip 
strength $u_\mathrm{t}$ for the simplified model. 
The grey solid line shows the coherent quantum conductance at zero 
temperature. 
The black line is the coherent conductance at $T=\unit[300]{mK}$, obtained 
from a convolution of the energy-dependent zero-temperature conductance 
with the derivative of the Fermi distribution, while the light blue and 
dark blue (dashed) lines represent the results obtained from our analysis 
based on classical trajectories (see Sec.\ \ref{sec:trajectories} for details). 
The main features of the zero-temperature quantum conductance are the same 
in the realistic simulation and in the simplified model. 

For large tip strength the conductance decreases in quantized steps 
of height $2 \times (2e^2/h)$ as in the realistic model. 
For weaker tip strength, large scale 
oscillations with an amplitude of about $6 \times (2e^2/h)$ dominate the 
the tip-dependence of the conductance. 
Strikingly, and even though the tip potential represents a repulsive 
obstacle for the electrons, at some tip strengths the conductance even exceeds 
that of the unperturbed cavity. 
Superposed with this secular behavior, the zero-temperature conductance 
exhibits ballistic conductance fluctuations with an amplitude of 
about $2 \times (2e^2/h)$. 
Those conductance fluctuations are expected to decay with temperature on 
a scale that corresponds to the correlation energy of the fluctuations. 
In classically chaotic systems the latter can be estimated within a 
semi-classical approach 
\cite{bluemel1988,jalabert1990} to be of the order of 
$\hbar/\tau=\hbar v_\mathrm{F}/\langle L \rangle$, where $\tau$ is the 
average time an electron spends in the cavity, 
$\langle L \rangle$ the average length of classical trajectories through the 
cavity and $v_\mathrm{F}$ the Fermi velocity. Assuming 
$\langle L \rangle\approx W/\pi R_0$ and using the chaotic prediction with 
the parameters of our system yields a temperature scale of 
about \unit[75]{mK}. 
Consistent with such an estimate, we find numerically that the fluctuations 
are indeed suppressed at the temperature of 
$T=\unit[300]{mK}$ that is used in the experiment, thereby confirming their 
quantum origin. In contrast, the larger scale oscillations remain robust, 
pointing to a different mechanism determining their occurrence. We show in the 
following section that they are of classical origin.  

\section{Analysis in terms of classical trajectories}\label{sec:trajectories}

In the ballistic regime, the conductance in the classical limit of 
$\hbar \to 0$ (where quantum interference is suppressed) can be expressed in 
terms of classical trajectories traveling between entrance and exit of the 
structure as 
\cite{baranger1991,jalabert_scholarpedia}
\begin{equation}\label{eq:gclassgen}
g_\mathrm{class}=\frac{m v_0 W}{\hbar \pi}\mathcal{T}\, ,
\end{equation}  
up to a constant whose value is not accessible by the semiclassical 
approach leading to Eq.\ \eqref{eq:gclassgen} and which we will ignore. 
The quantities $m$ and $v_0$ are, respectively, the mass and initial 
velocity of the electrons at the Fermi energy. The factor 
$m v_0 W/(\hbar \pi)$ 
stands for the incoming electron flux and its integer part is the number of 
propagating channels in the leads. 
For the case of GaAs, we use $m=0.067\, m_0$, with $m_0$ the free 
electron mass. The transmission probability is given by
\begin{equation}\label{eq:transmissionprob}
\mathcal{T}=\frac{1}{2}\int_{-\pi/2}^{\pi/2}\mathrm{d}\theta\, \cos\theta
\int_{-W/2}^{W/2}\mathrm{d}y\, f(y,\theta)\, ,
\end{equation} 
where $f(y,\theta)=1\, (0)$ for transmitted (reflected) trajectories that 
enter the cavity at a cross-section in the left contact at $y$ and with 
momentum direction characterized by the angle $\theta$ with respect to the 
$x$-axis. \cite{symmetries}
In quantum billiards, where the electrostatic potential is either zero 
or infinity, the trajectories depend on the geometry but not on the 
electron energy. 
Therefore $\mathcal{T}$ is independent of the energy and can be obtained 
from the asymptotic values of the quantum conductance in the limit of 
infinite energy.
In the case under study, we do not have a quantum billiard, due to the smooth 
character of the tip potential, and therefore $\mathcal{T}$ is 
energy-dependent. 

\begin{figure*}
\centerline{\includegraphics[width=0.85\textwidth]{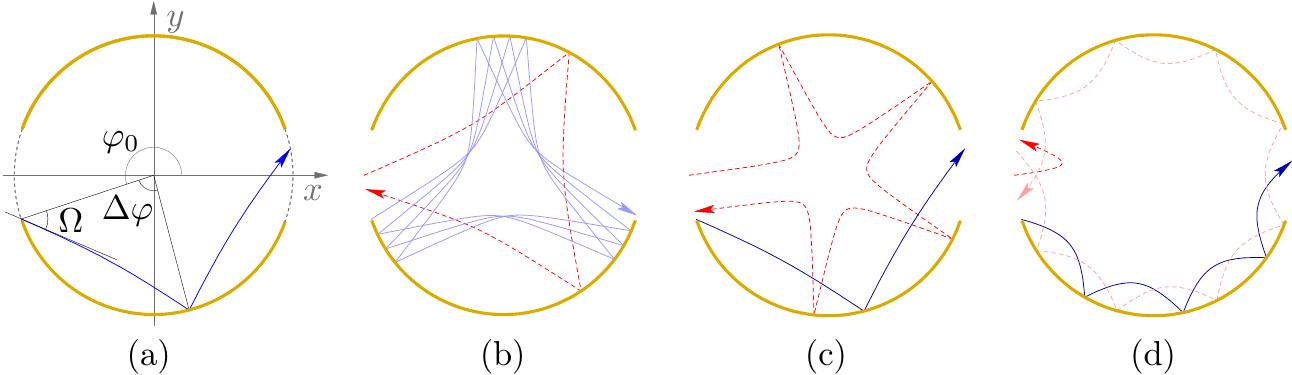}}
\caption{\label{fig-tra} (Color online) (a) Definition of the coordinate system and 
the angles used to characterize individual trajectories. $\varphi_0$ is 
the angle of the initial position with respect to the $x$-axis, 
$\Omega$ denotes the angle between the initial momentum of the trajectory 
and the diameter of the circle, and $\Delta\varphi$ the 
angular distance between two subsequent collisions with the cavity wall.
Two examples of transmitted (blue solid) and reflected (red dashed) classical trajectories 
are shown in (b) for tip strength $u_\mathrm{t}=0.03$ and in (c) 
for $u_\mathrm{t}=0.06$, three trajectories are shown in (d) for a rather 
strong tip characterized by $u_\mathrm{t}=0.43$.
}
\end{figure*}  
Equation \eqref{eq:gclassgen} for a ballistic system is the equivalent 
of the Drude 
conductance for the disordered case. In ballistic structures, the smooth 
disorder only weakly affects trajectories that are considerably shorter 
than the transport mean free path. Therefore, if the properties under study 
are dominated by the contribution of short trajectories, it is possible to 
ignore the disorder altogether and to include only the trajectories of the 
clean (disorder-free) geometry.

The function $f(y,\theta)$ is in general difficult to determine, and the 
integrals in Eq.\ \eqref{eq:transmissionprob} are typically calculated by 
randomly sampling the initial conditions.
However, the case of a hard-wall circular cavity with a central tip 
represented by the potential \eqref{pot_minimal} is quite simple 
due to the conservation of angular momentum within the cavity and the 
availability of an analytical expression describing the trajectories.

A trajectory is determined by the initial conditions of a particle in the 
left contact, described by its energy, position, and momentum orientation. 
The analytical expression of trajectories in the potential \eqref{pot_minimal} 
and specular reflection laws allow us to determine the subsequent points 
where the electron hits the circle of radius $R_0$ representing the cavity 
wall (see inset of Fig.\ \ref{fig:conductance} (b) and Fig.\ \ref{fig-tra}). 
As soon as such a point lies in one of the contacts 
between the cavity and the leads (i.e. $|y|<W/2$), the trajectory is 
completed and contributes to the transmission (reflection) probability if it 
reaches the right (left) exit.   
Figure \ref{fig-tra} shows a few examples of transmitted (blue solid) and 
reflected (red dashed) trajectories at different tip strength, 
which are discussed below. 

The description of a trajectory can be reduced to an equidistant series of 
position angles $\varphi_{n}=\varphi_{0}\pm n\Delta\varphi$ where the 
particle hits the cavity edge, for trajectories turning 
counterclockwise (clockwise) around the cavity center. 
We denote by $\varphi_{0}$ the initial angle with respect to the $x$-axis describing 
the starting point in the left contact (see Fig.\ \ref{fig-tra} (a) for a sketch), 
which is taken on the dashed line at radius $r=R_0$, \cite{startingonacircle} 
and $\Delta\varphi$ the (positive) rotation angle, which is a characteristic 
parameter of the trajectory. 

The classical results shown in Fig.\ \ref{fig:conductance} (b) are based on the 
application of Eq.\ \eqref{eq:gclassgen} in the framework of the simplified 
model. 
For each value of $u_\mathrm{t}$, we sampled a finite ensemble  of 
trajectories $\mathcal{M}$ corresponding to different initial conditions 
(at fixed energy) in the left contact. 
The prescription \cite{baranger1991} to calculate the classical transmission  
using eqs.\ \eqref{eq:gclassgen} and \eqref{eq:transmissionprob} is 
implemented adopting homogeneously distributed transverse starting points and, 
since $\mathrm{d}\theta\, \cos\theta= \mathrm{d}(\sin\theta)$, 
a sampling of initial angles $\theta$ such that the values of $\sin\theta$ 
are equally spaced. Such a sampling yields the expression 
\begin{equation}
g_\mathrm{class}=
\frac{m v_0 W}{\hbar \pi}
\frac{\#\mathcal{M}_\mathrm{T}}{\#\mathcal{M}}
\label{classicalconductance}
\end{equation}
for the classical conductance in terms of the cardinalities (denoted by $\#$) 
of the subset of transmitted trajectories $\mathcal{M}_\mathrm{T}$ and that of 
the total set of sampled trajectories 
$\mathcal{M}=\mathcal{M}_\mathrm{T}\cup\mathcal{M}_\mathrm{R}$ 
($\mathcal{M}_\mathrm{R}$ denotes the subset of reflected trajectories).

The classical conductance (light and dark blue dashed lines in 
Fig.\ \ref{fig:conductance} (b)) has only a very weak temperature dependence 
(at least up to $T=\unit[300]{mK}$). 
Its behavior is remarkably close to the finite-temperature 
quantum conductance in the regime of not too strong tip strength. 
The small offset of our classical conductance with respect to the quantum 
can be attributed to the ignored constant in Eq.\ \eqref{eq:gclassgen}. 
In particular, the classical results clearly exhibit the tip-strength 
dependent oscillations. 
Their ability to describe the behavior of the numerically calculated quantum 
conductance demonstrates that the large scale oscillations are of 
classical origin. 
In contrast, the conductance fluctuations and the conductance 
quantization for strong tip strength are quantum effects and therefore 
not present in the classical results. 
We conclude that the experimentally observed behavior at 
$T=\unit[300]{mK}$ is very well described by the classical 
treatment of our simplified model, 
except for the conductance quantization at large tip strength. 

In the sequel of the paper, we present a detailed analysis in order to 
understand why the rich variety of classical trajectories (some of them shown 
in Fig.\ \ref{fig-tra}) results in the simple structure of $g_\mathrm{class}$ 
presented in Fig.\ \ref{fig:conductance}. 
The properties of the classical trajectories through the sample and their 
dependence on tip strength will be analyzed towards an understanding of the 
mechanism that leads to the large conductance oscillations.

\begin{figure*}
\centerline{\includegraphics[width=0.6\linewidth]{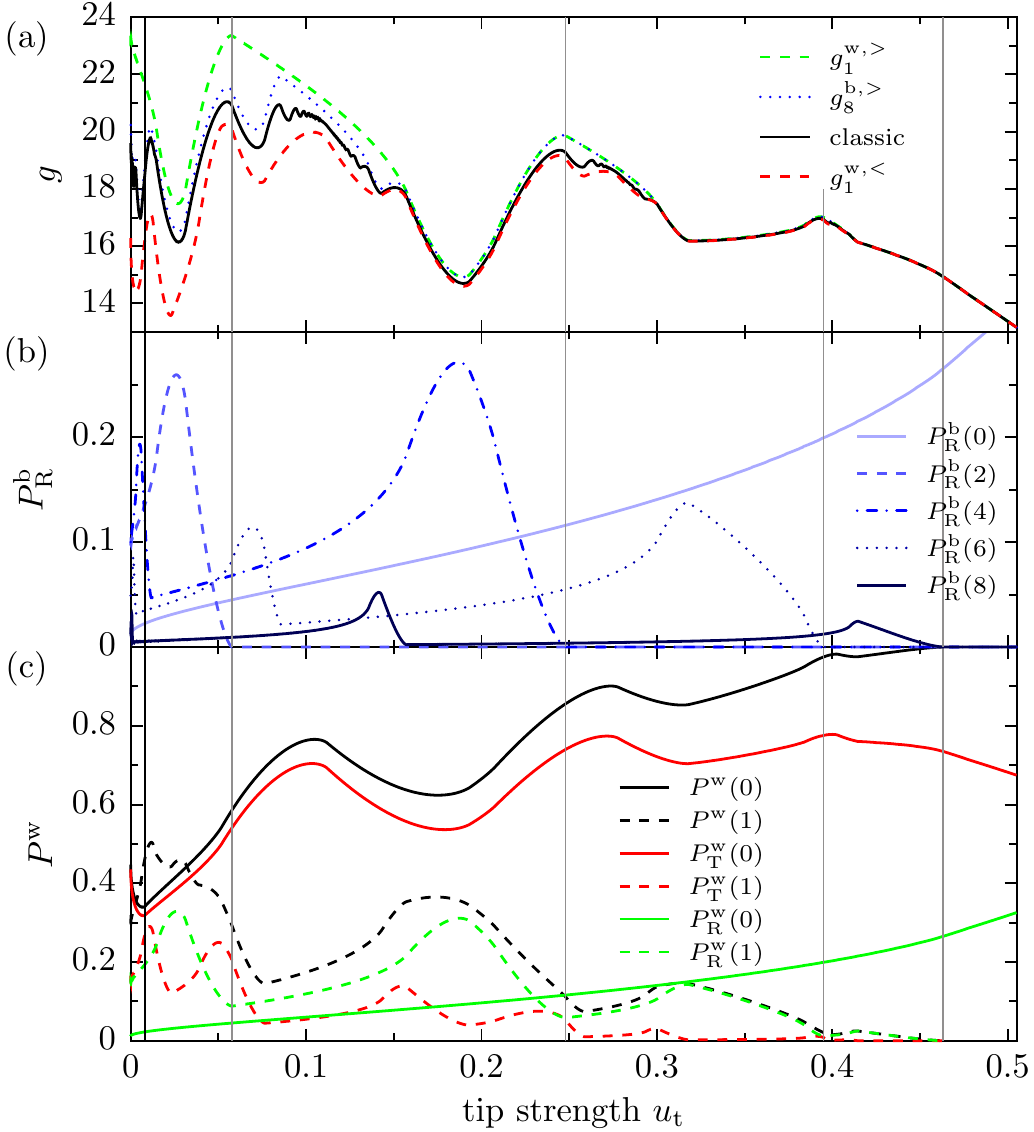}}
\caption{\label{fig:probabilities} (Color online) 
Influence of tip strength on 
(a) the classical conductance and conductance bounds obtained with
$g^{\mathrm{b},>}_8$, $g^{\mathrm{w},>}_1$ and $g^{\mathrm{w},<}_1$.
(b) Probabilities $P^\mathrm{b}_\mathrm{R}$ of an 
electron to be reflected after 
$b_s=\{0,2,4,6,8\}$ bounces with the cavity wall. 
(c) Probabilities $P^\mathrm{w}$ of trajectories 
without (solid) and one winding (dashed) 
around the center of the cavity (black) and its reflected 
$P^\mathrm{w}_\mathrm{R}$ (green, light) and transmitted 
$P^\mathrm{w}_\mathrm{T}$ (red, medium) parts. 
The leftmost vertical line indicates the tip strength above which 
transmitted trajectories without bounces are impossible and 
the following vertical lines indicate values above which reflected 
trajectories with $2$, $4$, $6$, and $8$ bounces are impossible.      
}
\end{figure*}
\section{Mechanism leading to conductance oscillations}\label{sec:mechanism}

In order to understand the origin of the large-scale conductance oscillations 
at low tip strength we investigate the dependence of the contributing 
trajectories on the strength of the tip potential. 
We characterize each trajectory $s$ by two parameters: 
The number of bounces at the cavity edge $b_s$, and the number 
of windings around the cavity center $w_s$. 
We distinguish transmitted ($\alpha=\mathrm{T}$) and reflected 
($\alpha=\mathrm{R}$) trajectories.

The probability that a trajectory is transmitted or reflected after 
$b$ bounces with the cavity edge can be calculated as  
\begin{equation}
P^\mathrm{b}_\alpha(b)=\frac{\sum_{s\in \mathcal{M}_\alpha} 
\delta_{b,b_s}}{\#\mathcal{M}}
 \label{P-col}
\end{equation}
with $\alpha=\mathrm{T}$ and $\mathrm{R}$, respectively.
The calculation of the probabilities of having $w$ windings 
$P^\mathrm{w}_\alpha(w)$ is analogous to Eq.\ \eqref{P-col}. 
We also introduce the probabilities 
$P^\mathrm{b}(b)=\sum_\alpha P^\mathrm{b}_\alpha(b)$ 
and $P^\mathrm{w}(w)=\sum_\alpha P^\mathrm{w}_\alpha(w)$ of 
having a trajectory with $b$ bounces and $w$ windings, 
respectively. 

Due to the wide openings of the cavity, most of the features of the problem 
at hand can be explained by only considering 
relatively short trajectories having few bounces with the cavity wall 
($b_s\leq 8$) and very few windings around the cavity center 
($w_s\leq 1$). 
This assumption is confirmed by the comparison shown in 
Fig.\ \ref{fig:probabilities} (a) between the full classical conductance 
given by Eq.\ \eqref{classicalconductance} (black solid line) 
and restricted sums over trajectories with bounded numbers of 
bounces (dotted line) or windings (dashed lines).
In particular, restricting the reflected trajectories leads to 
upper bounds of the conductance 
\begin{eqnarray}
g^{\mathrm{b},>}_{b_\mathrm{max}}&=&\frac{m v_0 W}{\hbar \pi}
\left( 1-\sum_{b=0}^{b_\mathrm{max}} P^\mathrm{b}_\mathrm{R}(b) \right)\\
  g^{\mathrm{w},>}_{w_\mathrm{max}}&=&\frac{m v_0 W}{\hbar \pi}
\left( 1-\sum_{w=0}^{w_\mathrm{max}} P^\mathrm{w}_\mathrm{R}(w) \right),
\end{eqnarray}
while restricting the transmitted trajectories provides lower bounds 
\begin{equation}
g^{\mathrm{b},<}_{b_\mathrm{max}}=\sum_{b=0}^{b_\mathrm{max}} g^{\mathrm{b}}(b) 
\quad \mathrm{and} \quad
g^{\mathrm{w},<}_{w_\mathrm{max}}=\sum_{w=0}^{w_\mathrm{max}} g^{\mathrm{w}}(w) 
\end{equation}
where
\begin{equation}
g^{\mathrm{b}}(b)=\frac{m v_0 W}{\hbar \pi}
 P^\mathrm{b}_{\mathrm{T}}(b)
 \quad \mathrm{and} \quad
g^{\mathrm{w}}(w)=\frac{m v_0 W}{\hbar \pi}
 P^\mathrm{w}_{\mathrm{T}}(w)
\end{equation}
are the conductance contributions of trajectories with $b$ bounces and $w$ windings, respectively. The bounds $g^{\mathrm{b},>}_8 $, 
$g^{\mathrm{w},>}_1$ and $g^{\mathrm{w},<}_1$ already contain the most 
important features of the conductance. They approach each other 
and the classical limit with increasing tip strength, 
indicating that the effect of long trajectories is suppressed 
by strong tips. 

The probabilities for reflected trajectories 
$P^\mathrm{b}_\mathrm{R}$ with an even number of 
bounces $b_s=\{0,2,4,6,8\}$ are presented in 
Fig.\ \ref{fig:probabilities} (b). \cite{noreflectionwithoddbounces} 
The case of $b=0$ concerns the direct 
reflection from the tip without touching the cavity edge, as e.g.\ the 
dashed red trajectory in Fig.\ \ref{fig-tra} (d).
The weight of this class of trajectories $P^\mathrm{b}_\mathrm{R}(0)$ 
grows with tip strength due to the increasing   
bending of the trajectories. For the same reason, the probability to be 
transmitted increases for trajectories that enter the cavity, hence 
reflected trajectories with a finite number of bounces are suppressed. 
The tip strengths where reflected trajectories with 
$b_s = \{2,4,6\}$ disappear (vertical lines 
in Fig.\ \ref{fig:probabilities}) are very close to the conductance maxima. 
We conclude that \textit{the disappearance of such a category of 
reflected trajectories is related to a maximum in the transmission}. 
This is an important element of the mechanism underlying the observed 
large-scale conductance oscillations. Consistent with our previous findings,
trajectories with $b_s\geq 8$ have little effect on the behavior of 
the conductance.

In order to illustrate the special role of the tip in selecting certain 
classes of transmitted or reflected trajectories, we discuss the example of 
trajectories with $b_s=2$ (dashed red curve in Fig.\ \ref{fig-tra} (b)). 
For the reflected trajectories $\Delta \varphi$ must satisfy 
$ 2\pi- 2\varphi_{\mathrm{op}} <3 \Delta \varphi <
2\pi+ 2\varphi_{\mathrm{op}}$,
where $\varphi_{\mathrm{op}}=\arcsin(W/2R_0)$ is the opening angle of 
the contacts (see inset of Fig.\ \ref{fig:conductance} (b)). 
With increasing tip strength $\Delta\varphi$ decreases and beyond a certain 
value the reflected trajectories with $b_s=2$ no longer exist.
Concomitantly, the transmitted trajectories with $b_s=1$ where 
$\pi-2\varphi_{\mathrm{op}} <2 \Delta \varphi <\pi+2\varphi_{\mathrm{op}}$
gain importance. The blue solid line in Fig.\ \ref{fig-tra} (c) is an example of such 
a trajectory at $u_\mathrm{t}=0.06$. 
The structure of the current density for this regime of tip strength shown 
in Fig.\ \ref{fig:localcurrent} demonstrates that the quantum current flow is 
closely related to the shape of those transmitted trajectories.
Reflection after a full winding is now only possible for trajectories with at 
least four bounces with the cavity 
wall (see red dashed line in Fig.\ \ref{fig-tra} (c)).      
This scenario is confirmed by the probabilities for zero and one windings 
$P^\mathrm{w}(0)$ and $P^\mathrm{w}(1)$ shown in Fig.\ \ref{fig:probabilities} (c).   
The number of transmitted trajectories with no winding around the center 
$P^\mathrm{w}_\mathrm{T}(0)$ is increasing while the probability to find a 
reflected trajectory with one winding $P^\mathrm{w}_\mathrm{R}(1)$ assumes 
minima close to the vertical lines.
Hence, the decrease of the weight of longer trajectories (a very long one 
is shown by the blue solid curve in Fig.\ \ref{fig-tra} (b)) with tip strength, and 
\textit{the alternation in the suppression of reflected and transmitted 
families explains the oscillations of the conductance}. 

In general, the reduced $\Delta \varphi$ increases the probabilities to be 
either reflected without collision or to be transmitted after an increasing 
number of bounces but without a full winding around the cavity center. 
This is illustrated in Fig.\ \ref{fig-tra} (d) showing three trajectories 
at tip strength $u_\mathrm{t}=0.43$, where reflected trajectories with one 
winding and $b_s = 8$ are still possible (see dotted red line). 
Once the tip strength imposes $\Delta \varphi < 2\varphi_{\mathrm{op}}$, 
the bending of the trajectories becomes so strong that all 
trajectories are either immediately reflected, as the dashed red curve in 
Fig.\ \ref{fig-tra} (d)), or transmitted after $b_s\geq 3$ collisions, 
as the blue solid curve in Fig.\ \ref{fig-tra} (d). 
Trajectories with a non-zero winding number become impossible and 
$P^\mathrm{w}(0)=1$, such that there is no conductance maximum at the tip 
strength where reflection after $8$ bounces becomes 
impossible. 

With a further increase of tip strength, the increase of 
direct reflection $P^\mathrm{w}_\mathrm{R}(0)$ continues, implying a  
decrease of the transmissions $P^\mathrm{w}_\mathrm{T}(0)$, and the 
conductance decreases monotonically with tip strength, reaching 
zero when the tip-induced depletion is so strong that electrons cannot enter 
the cavity any more.
Beyond the behavior accessible by our classical analysis, the
quantum conductance exhibits in this regime quantized 
values that are due to the two parallel quantum wires which 
are formed between the cavity wall and the tip.    

The initial decrease and the first minimum of the conductance at 
$u_\mathrm{t}\approx 0.01$ is due to the suppression of direct transmission  
without ever touching the cavity edge. 
Above the tip strength where direct trajectories cease to 
exist (leftmost vertical line in Fig.\ \ref{fig:probabilities}),
the tip is no longer perturbative in the quantum mechanical 
sense. \cite{firstorderperturbation}
   
\begin{figure}
\centerline{\includegraphics[width=\linewidth]{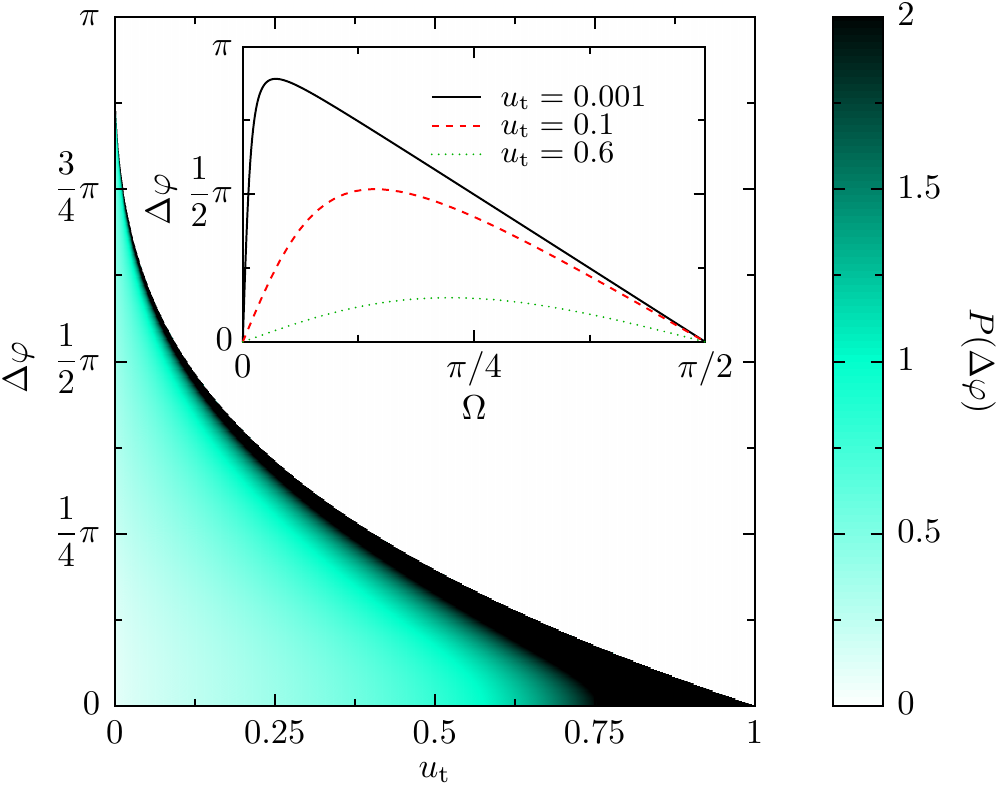}}
\caption{\label{fig:delta-phi} 
(Color online) Density plot of the probability distribution 
$P(\Delta \varphi)$ as a function of $u_\mathrm{t}$. 
Inset: Dependence of $\Delta\varphi$ on the injection angle 
$\Omega$, for different values of $u_\mathrm{t}$.}
\end{figure}
\section{Simplified analysis of the classical trajectories}\label{sec:simpleanalysis}

The previous statistical analysis shows that a reduced subensemble 
of classical trajectories suffices to 
explain the dependence of the conductance on tip strength, and 
points to the crucial importance of the parameter $\Delta\varphi$ 
that characterizes the angular distance between subsequent points 
where the trajectory hits the cavity boundary.
In this section we investigate the distribution of $\Delta\varphi$ 
at a given tip strength. An analytically tractable analysis based 
on the dominating
trajectories provides the essence of the mechanism underlying 
the classical conductance oscillations, as well as the recipe to 
predict the main conductance maxima and minima. 
This is particularly useful since each of the previously 
analyzed subensembles of classical 
trajectories is actually infinite and mixes very different 
behaviors.

Figure \ref{fig:delta-phi} shows a density plot of the probability 
distribution 
of the rotation angle $P(\Delta \varphi)$ within our ensemble of 
trajectories for values of $u_{\rm t}$ between $0$ and $1$. 
For a given $u_{\rm t}$ the angle $\Delta\varphi$ is solely a 
function of the angle $\Omega$ (due to symmetry, $\Delta\varphi$ 
does not depend on $\varphi_0$). 
The $\Omega$-dependence exhibits a maximum value 
$\Delta\varphi_\mathrm{m}$ 
(see Fig.\ \ref{fig:delta-phi}, inset), which decreases with 
increasing tip strength. 
Figure \ref{fig:delta-phi} shows that the probability density 
is highly concentrated for values of $\Delta\varphi$ close to 
$\Delta\varphi_\mathrm{m}$, actually diverging when the maximum 
value of $\Delta\varphi_\mathrm{m}$ is approached from below.
This behavior is due to the long tails of the tip 
potential \eqref{pot_minimal} that lead to a smooth dependence of 
$\Delta\varphi$ on the injection angle $\Omega$ defined with respect 
to the cavity diameter (see Fig.\ \ref{fig-tra} (a)). 
Trajectories injected with $\Omega=0$ are reflected in a head-on 
collision with the tip and have $\Delta\varphi=0$. Then, $\Delta\varphi$ 
reaches its maximum value $\Delta \varphi_{\rm m}$ at an intermediate value of 
$\Omega$, and decreases upon a further increase of $\Omega$ back to $0$ 
when $\Omega=\pi/2$. Examples of this behavior are shown in 
the inset of Fig.\ \ref{fig:delta-phi}. The flat maximum occurring in the 
dependence of $\Delta \varphi$ on $\Omega$ leads to the divergence of 
$P(\Delta \varphi)$ at the maximum value $\Delta \varphi_{\rm m}$. 

\begin{figure}
\centerline{\includegraphics[width=\linewidth]{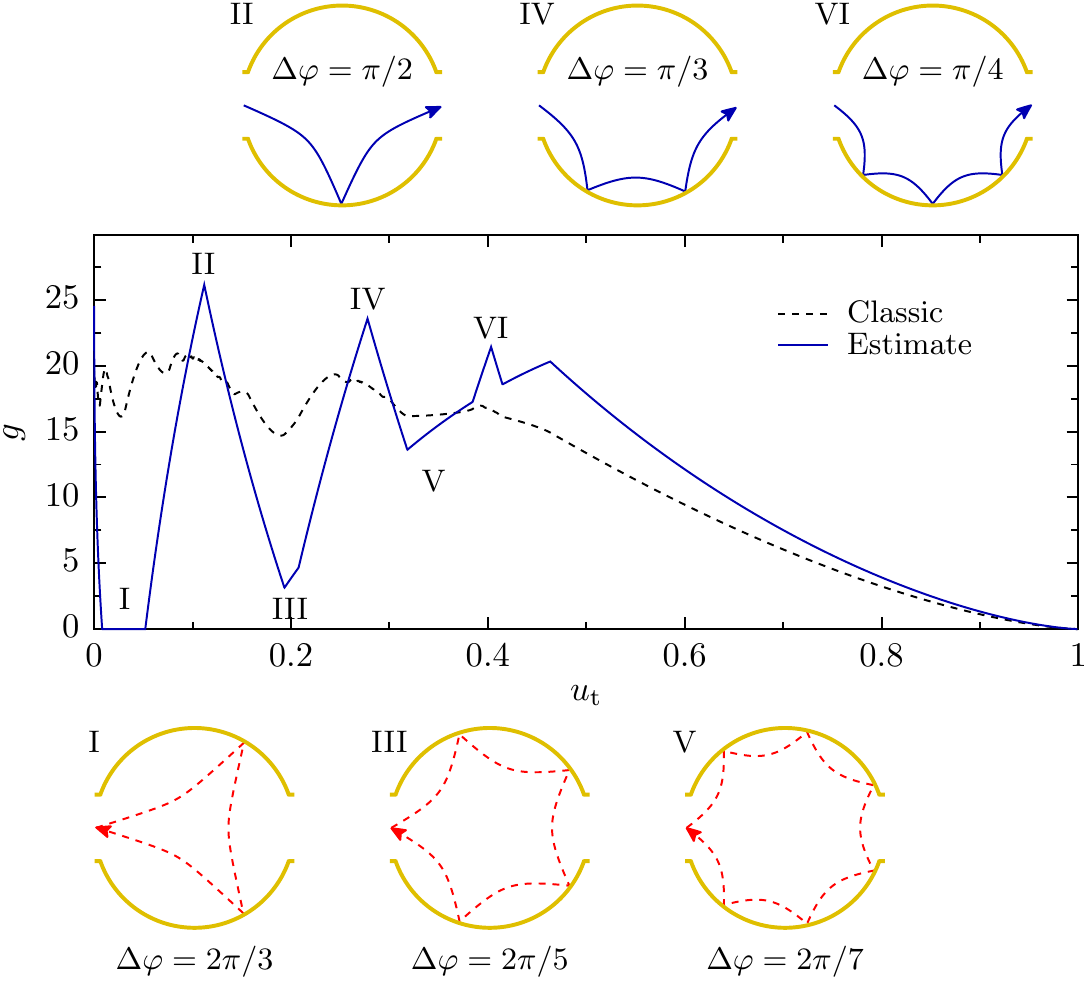}}
\caption{\label{fig:cond_estimate} 
(Color online) 
Conductance estimates \eqref{eq:estimate_large} and 
\eqref{eq:estimate_small}
based on the shortest trajectories with maximum $\Delta\varphi$ 
(blue solid line). 
The classical conductance (black dashed line) is shown for 
comparison. Sketches of the dominant classes of short transmitted 
and reflected trajectories in 
the regions of conductance maxima and minima are shown above 
and below the plot, respectively. 
}
\end{figure}
The resulting dominance of trajectories with $\Delta \varphi$ close to the 
maximum value $\Delta \varphi_{\rm m}$ becomes more pronounced with increasing 
values of $u_\mathrm{t}$ (and decreasing $\Delta \varphi_{\rm m}$), 
and \textit{motivates the characterization of the ensemble of classical 
trajectories at a given tip strength by the dominant angular distance
between bounces $\Delta \varphi_\mathrm{m}$}. 
We restrict in the sequel the analysis to only those trajectories having 
$\Delta \varphi = \Delta\varphi_\mathrm{m}$.

In the regime where the tip is so strong that 
$\Delta \varphi_\mathrm{m} < 2\varphi_\mathrm{op}$, all trajectories are 
reflected immediately without ever hitting the cavity wall, 
with the exception of those that start at a position that is separated by 
less than $\Delta \varphi_\mathrm{m}$ from the edge of the opening 
$\varphi_\mathrm{op}$. The latter trajectories are transmitted to the other 
lead after a number of bounces, but without winding around the center. Their 
proportion is given by $\Delta \varphi_\mathrm{m}/2\varphi_\mathrm{op}$ such 
that the classical conductance is estimated as 
\begin{equation}\label{eq:estimate_large}
g_\mathrm{class}\approx\frac{m v_0 W}{\hbar \pi}
\frac{\Delta\varphi_\mathrm{m}}{2\varphi_\mathrm{op}}\, .
\end{equation}
The decrease with $u_\mathrm{t}$ is a consequence of the 
increasing number of directly reflected trajectories when 
$\Delta \varphi_\mathrm{m}$ is suppressed by an increasing tip strength. 

For weaker tip strength, when 
$\Delta \varphi_\mathrm{m} > 2\varphi_\mathrm{op}$, the distance between 
bounces is larger than the openings of the cavity, and a direct 
reflection is impossible. 
Due to the relatively large openings of the cavity, 
longer trajectories are of minor importance, 
especially for large $u_\mathrm{t}$. For simplicity, we neglect them and 
concentrate on trajectories with no winding and transmitted 
after $b$ bounces. 
For such trajectories, the total rotation angle  
$(b+1)\Delta\varphi$ lies in the interval 
$\left[\pi-2\varphi_\mathrm{op},\pi+2\varphi_\mathrm{op}\right]$.
The evaluation of the share of initial positions in the left contact that lead 
to such a trajectory yields the estimate 
\begin{equation}\label{eq:estimate_small}
g^\mathrm{b}(b)\approx\frac{m v_0 W}{\hbar \pi}
\left(1-\frac{\left|\pi-(b+1)\Delta\varphi_\mathrm{m}\right|}
{2\varphi_\mathrm{op}}\right)
\end{equation}
for their contribution to the conductance. Thus, maxima of the total 
conductance, obtained by summing over $b$, 
can be expected when $\Delta\varphi_\mathrm{m}=\pi/(b+1)$. 
They are of triangular shape with a width $4\varphi_\mathrm{op}/(b+1)$. 

For the simplified model with the tip potential \eqref{pot_minimal}, the above 
criterion for transmission maxima translates into conductance 
peaks at $u_\mathrm{t}\approx \{0, 0.111, 0.278, 0.403 \}$ 
(blue solid curve in Fig.\ \ref{fig:cond_estimate}). 
Close to these values maxima arise in the probability of having a 
transmitted trajectory with zero winding 
(red line in Fig.\ \ref{fig:probabilities} (c)).
The estimates of the conductance \eqref{eq:estimate_large} and 
\eqref{eq:estimate_small} that are based on the shortest transmitted 
trajectories are represented by the blue solid line in 
Fig.\ \ref{fig:cond_estimate}  
that can be compared to the classical conductance (black dashed line). 
The classes of short transmitted trajectories that dominate in the 
vicinity of the conductance peaks are sketched above the plot.
The qualitative agreement of the peak structure confirms the 
dominance of those short trajectories. However, at small tip strength 
$u_{\rm t}$ the quantitative comparison becomes rather poor, pointing 
to the increased importance of longer trajectories 
and a broader distribution of the values of $\Delta\varphi$.

The geometric argument can be extended to longer trajectories. 
The class of trajectories that are reflected after one winding around the 
cavity lead to a minimum in conductance at the values
$\Delta \varphi_{\rm m}= 2\pi/(b +1)$ with the width 
$4\varphi_\mathrm{op}/(b+1)$, for even $b$. 
The corresponding conductance minima are expected at 
$u_{\rm t}\approx \{  0.031, 0.199, 0.345, 0.425\}$. 
These values are close to the minima in the classical conductance 
(Fig.\ \ref{fig:cond_estimate}), as well as to the maxima in the probability 
of having reflected trajectories with one winding   
(green dashed line in Fig.\ \ref{fig:probabilities} (c)). 
Also the probabilities $P^\mathrm{b}_R(b)$ of being reflected after 
$b$ bounces in Fig.\ \ref{fig:probabilities} (b) assume maxima close to 
the corresponding values of $u_{\rm t}$. The dominant classes 
of short reflected trajectories close to the conductance minima are sketched 
below the graph in Fig.\ \ref{fig:cond_estimate}.

The long range tip potential is a crucial 
ingredient for the mechanism presented above since it leads to the 
dominance of trajectories at $\Delta \varphi_\mathrm{m}$. 
In contrast, when the tip is modeled by a hard wall disc, 
the dependence of $\Delta\varphi$ on the injection angle has a cusp 
at the maximum. The resulting probability density does 
not exhibit a significant preferential value, the behavior of very 
different values of $\Delta\varphi$ is mixed, 
and the conductance oscillations are smeared, consistent 
with our numerical observations. The role of the 
steepness of the cavity potential is very different as it does not affect
significantly the specular reflection at the cavity edges. Thus, assuming hard wall boundaries for the cavity is a very good approximation, at least 
within the present study.
   
\begin{figure}
\centerline{\includegraphics[width=\linewidth]{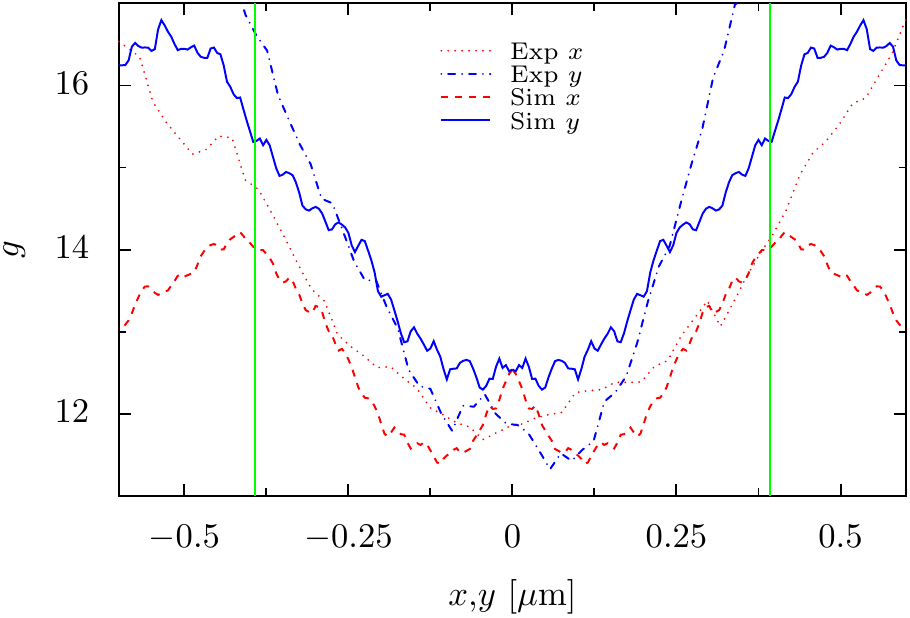}}
\caption{\label{fig:offcenter}
Conductance cuts as a function of tip position in $x$ (red) and $y$ (blue) 
directions away from the cavity center. Solid and dashed lines 
represent the realistic numerical simulation at 
$u_\mathrm{t}=0.0315$ and $T=\unit[300]{mK}$, dash-dotted and 
dotted lines the experimental conductance at 
$V_\mathrm{Tip}=\unit[-3]{V}$.   
The green vertical lines represent the rough estimate $y/R_0\approx \pi/12$
for the size of the central spot in Fig.\ \ref{fig:experiment} (b) (see text).
}
\end{figure}
\section{Off-center tip}\label{sec:off-center}

The experimental SGM scans of Fig.\ \ref{fig:experiment} (b-c) show that the 
non-monotonic behavior of the conductance as a function of tip strength 
extends to a well-defined region around the center of the cavity.
It is important to understand the robustness of this surprising effect, since 
moving the tip off-center has the important consequence of breaking the 
integrability of the underlying classical dynamics.

In Fig.\ \ref{fig:offcenter} we present cuts of the experimental data of 
Fig.\ \ref{fig:experiment} (b) corresponding to the tip voltage 
$V_\mathrm{Tip}=\unit[-3]{V}$ for the two axis directions $x$ and $y$, 
together with the numerical simulation within the realistic model. 
The agreement between the two sets of data is quite good. The deviations 
obtained for tips approaching the border of the cavity are probably due to 
our approximation of superposing the tip-induced potential and the 
confinement potential. The small-scale structure in the numerical traces is
less pronounced in the experiment, where inelastic processes suppress
the contribution of long trajectories.

It is important to remark that the numerical calculations reproduce the width 
of the conductance dip observed in the cavity center and even the asymmetry 
of the central spot in Fig.\ \ref{fig:experiment} (b), indicating a faster 
recovery of high conductance values when moving away from the center in the 
$y$-direction, as compared with the $x$-direction. As in the previously 
studied case of a centered tip, the physical understanding of the spatial 
extension of the non-monotonic dependence of conductance on tip voltage 
requires the study of the simplified model and a semiclassical analysis. 
The numerics in the simplified model (not shown) is consistent with the 
experimental results and the spatial anisotropy. Within the simplified 
analysis in terms of classical trajectories we can make a rough estimate of
the size of the central spot in Fig.\ \ref{fig:experiment} (b).

The minimum conductance observed at $V_\mathrm{Tip}=\unit[-3]{V}$ 
is due to the dominance of reflected ``triangular" 
trajectories with $b=2$ bounces and $w=1$ winding 
(see Fig.\ \ref{fig:probabilities}) that are most prominent when 
$\Delta \varphi_\mathrm{m} \approx 2\pi/3$. The subsequent conductance 
maximum corresponds to the dominance of trajectories with $b=1$ and $w=0$. 
Those trajectories correspond to $\Delta \varphi_\mathrm{m} \approx \pi/2$ 
that are transmitted after a single bounce at the cavity edge, and that can 
pass above or below the tip. We assume that those short trajectories, 
although deformed, still determine the behavior when the tip is slightly
off-centered. For a displacement in the $y$-direction, 
the angular distance of the bounces seen from the tip 
position is modified to  $\Delta \varphi \approx \pi/2\pm y/R_0$. 
Assuming that 
the conductance oscillations are washed out when the resulting difference of 
$2y/R_0$ in $\Delta \varphi$ between trajectories above and below the tip 
reaches the difference $2\pi/3-\pi/2$ corresponding to 
conductance maximum and minimum, we get the estimate    
$y/R_0\approx \pi/12$ for the distance $y$ between the tip 
and the cavity center in transverse direction where 
the classical oscillations are expected to be smeared. 
This universal value, independent 
of the size of the cavity openings, is indicated by the green vertical lines 
in Fig.\ \ref{fig:offcenter}. It is consistent with the experimental 
findings (size of the central region of suppressed conductance 
in Fig.\ \ref{fig:experiment} (b)) and also with our numerical calculations. 

\section{Conclusions}\label{sec:conclusions}

We presented SGM experiments on a circular cavity in which an intriguing 
non-monotonic dependence of the conductance on the tip voltage is observed 
when the tip is placed near the center of the cavity. Our theoretical analysis 
explains the unexpected behavior and traces it back to classical electron 
dynamics in the device.

The numerical simulation of quantum transport through the experimental setup 
with a long-range tip potential in the center at the temperature of the 
experiment yielded large conductance 
oscillations as a function of tip strength that are 
in quantitative agreement with the 
experimental findings (see Fig.\ \ref{fig:conductance} (a)).
At low temperature, additional ballistic conductance fluctuations appear 
in the numerically calculated conductance. 

The quest for a physical understanding of the surprising non-monotonic 
behavior found in the experiments and in the realistic quantum simulations 
lead us to proceed in three steps. First, we simplified the model in order to 
have, at the same time, the quantum simulations exhibiting the observed 
behavior, and an analytically tractable classical dynamics. Second, we 
performed a statistical analysis of short classical trajectories that 
linked the conductance oscillation with the switches between groups of 
dominant trajectories. Third, we developed a simplified analysis that 
identified the specific features of the trajectories that result in the 
experimentally observed phenomenon.  

While the quantum conductance fluctuations are suppressed at the 
temperature of the experiment, the classical conductance is not 
significantly affected by temperature. The large scale conductance 
oscillations obtained in the classical limit reproduce the behavior 
of the quantum conductance at the experimental temperature, except for 
conductance steps that arise in the regime of very strong tips.  
We therefore conclude that the oscillations of the conductance as a 
function of tip strength and the resulting non-monotonic behavior of the 
conductance when the tip is in the center of the cavity is of classical 
origin.  

The statistical analysis of the ensemble of trajectories, 
possible in the simplified model system, shows that relatively short 
trajectories with few bounces with the 
cavity wall or few windings around the cavity center allow to understand the 
conductance oscillations. 
The evolution of the contributing trajectories with tip strength provides 
the mechanism for the observed conductance oscillations. 

Having observed that trajectories with a particular angular distance between 
bounces at the cavity wall dominate in the case of strong tip potentials, 
we showed that the restriction to this class of dominating 
trajectories leads to a basic understanding of the main conductance maxima.
We found that the long range character of the tip potential is a crucial 
ingredient for the classical mechanism. 
When the tip is modeled as a hard disc of increasing size, the 
experimentally observed non-monotonic behavior of the conductance cannot 
be reproduced.   

The classical analysis presented in this work is able to account for the 
conductance oscillations as a function of the tip strength for the case of 
centered tips, but also for the experimentally observed decay of these 
oscillations as the tip moves away from the cavity center.

It is remarkable that a simple modeling based on relatively short classical 
trajectories, completely ignoring disorder and electron-electron 
interaction, was capable of rendering the explanation of the measured 
conductance through a cavity within an SGM setup. 
Even if the modeling of ballistic transport is usually done for 
quantum billiards, the role of the smoothness of the electrostatic 
confinement potential has been discussed in the 
literature. \cite{ouchterlony99,marlow06}
In our work we have shown that the electrostatic confinement defining 
the structure seems to be sufficiently sharp as not giving rise to 
important departures from the hard-wall case. 
However, the smooth character of the electrostatic potential created by the 
tip imprints a crucial signature for the electric transport with an SGM 
setup.

The role of short trajectories in ballistic transport has been 
pointed out in the context of various experimental 
setups. \cite{jalabert_scholarpedia}
In particular, it has been shown \cite{jacquod04} that for classically 
chaotic cavities the deterministic sector of the phase space 
corresponding to short trajectories may give rise to conductance 
oscillations that are more pronounced than the conductance 
fluctuations stemming from the stochastic sector. 
The relative weight of the deterministic and stochastic sectors of 
phase space can be changed by varying the openings of the cavity or, 
within a fixed geometry, by the application of a magnetic field 
(as in experiments measuring shot noise in chaotic 
cavities \cite{oberholzer02}).

While the signature of classical trajectories in transport through circular 
billiards has been identified \cite{berry94,ishio95,jalabert_scholarpedia} 
by the Fourier transform of the energy and magnetic field-dependent 
conductance, our experimental and theoretical results provides a direct 
evidence of the almost exclusive role of a small class of trajectories.    
The correlation of magneto-resistance maxima with 
specific electron trajectories has been observed in the case of 
triangular \cite{linke97} and circular \cite{hackens02} cavities 
under a perpendicular magnetic field. The commensurability 
conditions between the cyclotron radius and the linear dimensions 
of the cavities result in magneto-conductance oscillations similar 
to the conductance oscillations that we study in this work. 
Our case presents the particularity that the curvature of the 
classical trajectories depends on the impinging angle, 
resulting in the domination of a given rotation angle between 
successive bounces with the walls of the cavity, leading to very large
conductance oscillations.

Experiments carried out in small cavities (diameter \unit[1.0-1.5]{$\mu$m}) 
do not show significant traces of the classical oscillations as compared to 
those of the large cavity of Fig.\ \ref{fig:experiment}. 
The small cavities do not have perfect circular symmetry, which might 
explain the different behavior.
Moreover, numerical simulations indicate a stronger difference between the 
Lorentzian and the simplified tip potential for the case of the smaller 
cavities and point to a modification of the mechanisms in the case where the 
width of the Lorentzian tip potential is not much smaller than the cavity. 

In the small cavities it is possible to achieve the regime in which the 
size of the tip-induced disk becomes comparable with the size of the cavity. 
\cite{kozikov14,steinacher15}
SGM measurements in this regime show fringes that extend to the center of the 
cavity and a conductance suppression with increasing negative tip voltage 
exhibiting steps. 
Such a behavior is consistent with the conductance quantization at large tip 
strength that can be observed in the result for the quantum conductance 
presented in Fig.\ \ref{fig:conductance}.

\acknowledgments

Financial support from the French National Research Agency 
ANR (Projects ANR-11-LABX-0058{\_}NIE, ANR-14-CE36-0007-01) and from the 
Swiss National Science Foundation SNSF is gratefully acknowledged.

\bibliography{references} 

\begin{thebibliography}{45}
\bibitem [{tip()}]{tipnotdependentonR}%
  \BibitemOpen
  \href@noop {} {}\bibinfo {note} {Such a choice affects the value of the
  dimensionless strength parameter $u_\mathrm{t}$, but it does not imply that
  the tip potential $U_\mathrm{S}(\vec{r})$ depends on the Fermi energy or the
  size of the cavity.}\BibitemShut {Stop}%
\bibitem [{nor()}]{noreflectionwithoddbounces}%
  \BibitemOpen
  \href@noop {} {}\bibinfo {note} {The symmetry of the system, with equal width
  openings exactly opposite to each other, makes reflection after an odd number
  of bounces impossible, such that $P^\mathrm{b}_\mathrm{R}(b)$ vanishes for
  all odd $b$, independent of the tip strength.}\BibitemShut {Stop}%
\end{thebibliography}

\end{document}